\documentclass{article}
\usepackage[pdftex]{graphicx}
\usepackage{amsmath}
\usepackage{amsthm}
\usepackage{netcod09}  

\begin{document}
\topmargin = 0mm

\itwtitle{Broadcasting in Time-Division Duplexing: A Random Linear Network Coding Approach}

\itwauthor{Daniel E. Lucani}
{
Massachusetts Institute of Technology\\
Cambridge, Massachusetts, 02139 \\
USA \\
Email: dlucani@mit.edu}

\itwsecondauthor{Muriel M\'edard}
{
Massachusetts Institute of Technology\\
Cambridge, Massachusetts, 02139 \\
USA \\
Email: medard@mit.edu}

\itwthirdauthor{Milica Stojanovic}
{
Northeastern University\\
Boston, Massachusetts, 02115\\
USA\\
Email: millitsa@mit.edu}



\itwmaketitle


\begin{itwabstract}
{\small We study random linear network coding for broadcasting in time division duplexing channels. We assume a packet erasure channel with nodes that cannot transmit and receive information simultaneously. The sender transmits coded data packets back-to-back before stopping to wait for the receivers to acknowledge the number of degrees of freedom, if any, that are required to decode correctly the information. We study the mean time to complete the transmission of a block of packets to all receivers. We also present a bound on the number of stops to wait for acknowledgement in order to complete transmission with probability at least $1-\epsilon$, for any $\epsilon>0$.     
We present analysis and numerical results showing that our scheme outperforms optimal scheduling policies for broadcast, in terms of the mean completion time. We provide a simple heuristic to compute the number of coded packets to be sent before stopping that achieves close to optimal performance with the advantage of a considerable reduction in the search time.}
\end{itwabstract}

\begin{itwpaper}

\itwsection{Introduction}

Reference \cite{lucaniInfocom09} considered the use of network coding in channels in which time division duplexing is necessary, i.e. when a node can only transmit or receive, but not both at the same time. This type of channel is usually called half-duplex, but we will use the term time division duplexing (TDD) to emphasize that the transmitters and receivers may not use the channel half of the time each, i.e. the amount of time allocated to transmit and receive may vary. Examples of TDD channels are infrared devices, and underwater acoustic modems. 
High latency channels, e.g. in satellite, and deep space communications, can also take advantage of these ideas.  

In particular, Reference \cite{lucaniInfocom09} studied the case of transmitting a block of $M$ data packets through a link using random linear network coding with the objective of minimizing the mean time to complete transmission of that block of packets. Reference \cite{lucaniICC09} extended the analysis for the problem of energy consumption of the scheme. Reference \cite{lucaniInfocom09} and \cite{lucaniICC09} showed that there exists, an optimal number of coded data packets to be transmitted back-to-back before stopping to wait for an acknowledgment (ACK), under the minimum time and minimum energy criterion, respectively. Reference \cite{lucaniICC09} showed that choosing the number of coded data packets to optimize mean completion time, as in \cite{lucaniInfocom09}, provides a very good trade-off between energy consumption and completion time.

We analyze the problem of broadcast under the TDD constrain and using a similar random linear network coding scheme to that in \cite{lucaniInfocom09}, i.e. we provide an extension of the scheme to the case of several receivers. We assume that the receivers are not allowed to cooperate in order to share their received coded packets in order to decode the information, which is to say that each receiver must be able to decode the information from the coded packets sent directly from the transmitter. We provide a bound to the number of stops to wait for ACKs in order to complete the transmission to all receivers with arbitrary high probability. Also, we study the mean completion time and energy of the broadcast scheme and compare it to optimal scheduling policies. Finally, we provide simple heuristics to determine the number of coded data packets to be transmitted back-to-back before stopping to wait for an ACK, with a considerable reduction in the computation time.  

	 The paper is organized as follows. In Section II, we outline the set up of the problem. We provide a proof for the bound on the number of stops to listen for ACK packets. In Section III, we study the mean completion time and heuristics to determine the number of coded packets. In Section IV, we present comparison schemes based on optimal scheduling policies. Section V presents numerical results for various broadcast scenarios. Conclusions are summarized in Section VI. 
	 
\itwsection{Random Network Coding for Broadcast in TDD channels}

	A sender wants to broadcast $M$ data packets at a given data rate $R$ [bps] to $N$ receivers as in Figure ~\ref{BroadcastNetwork.tag}. We assume an independent packet erasure channel for each of the receivers and that receivers cannot cooperate or share information. Nodes can transmit and receive, but not both at the same time. The sender uses random linear network coding \cite{ho06} to generate coded data packets. Each coded data packet contains a linear combination of the $M$ data packets of $n$ bits each, as well as the random coding coefficients used in the linear combination. Each coefficient is represented by $g$ bits. For encoding over a field size $q$, we have that $g = \log_2 q$ bits. A coded packet is preceded by an information header of size $h$. Thus, the total number of bits per packet is $h + n + gM$. Figure 2 in \cite{lucaniInfocom09} shows the structure of each coded packet. 

	The sender can transmit coded packets back-to-back before stopping to wait for an ACK packet from each receiver. Each ACK packet feeds back the number of degrees of freedom (dof), that are still required to decode successfully the $M$ data packets to a particular receiver. As in \cite{lucaniInfocom09}, we assume that the field size $q$ is large enough so that the expected number of successfully received packets at the receiver, in order to decode the original data packets, is approximately $M$. 


\begin{figure}[t]
\centering	
\includegraphics[height=1.5in,width=1.5in]{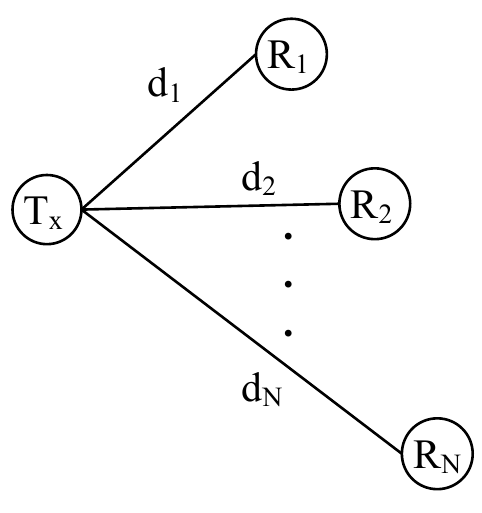}
\caption{Broadcast network.}
\label{BroadcastNetwork.tag}
\end{figure}


\begin{figure}[t]
\centering	
\includegraphics[height=1in,width=3.5in]{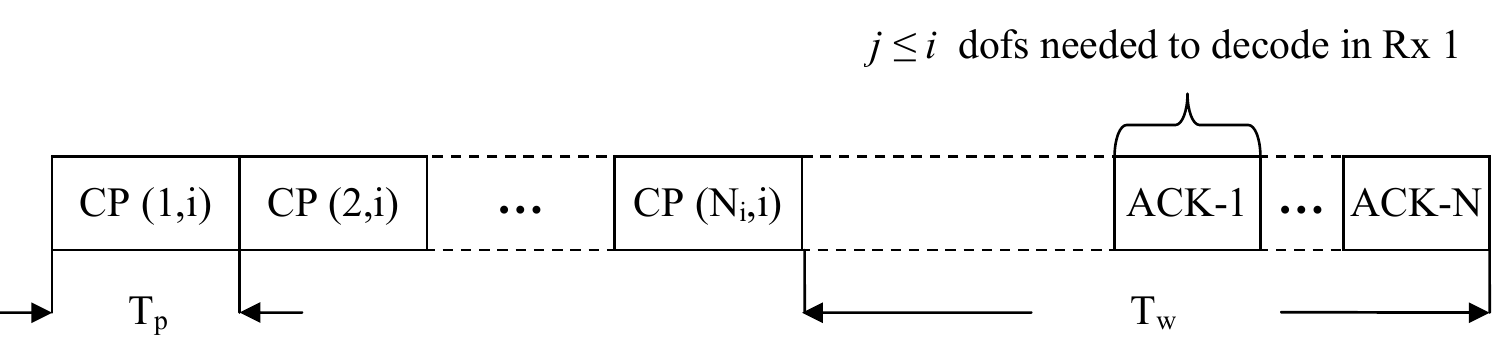}
\caption{Network coding TDD scheme.}
\label{Protocol.tag}
\end{figure}    

Transmission begins with $M$ information packets, which are encoded into $N_M \geq M$ random linear coded packets, and transmitted. If all $M$ packets are decoded successfully by all receivers, the process is completed. Otherwise, each ACK informs the transmitter how many dofs are missing, say $i_1, i_2, ..., i_N$ for receivers $1,2,...,N$, respectively. The transmitter then sends $N_i$ coded packets, where $i = \max _{j =1,2,...,N} i_j$. This process is repeated until all $M$ packets have been decoded successfully by all receivers. We are interested in the optimal number $N_{i}$ of coded packets to be transmitted back-to-back. 

Figure~\ref{Protocol.tag}, illustrates the time window allocated to the system to transmit $N_i$ coded packets. Each coded packet $CP(1,i)$, $CP(2,i)$, etc. is of duration $T_p$. The waiting time $T_w$ is chosen so as to accommodate the propagation delay and time to receive the ACKs from each receiver.

 \begin{figure}[t]
\centering	
\includegraphics[height=2.5in,width=2.5in,keepaspectratio]{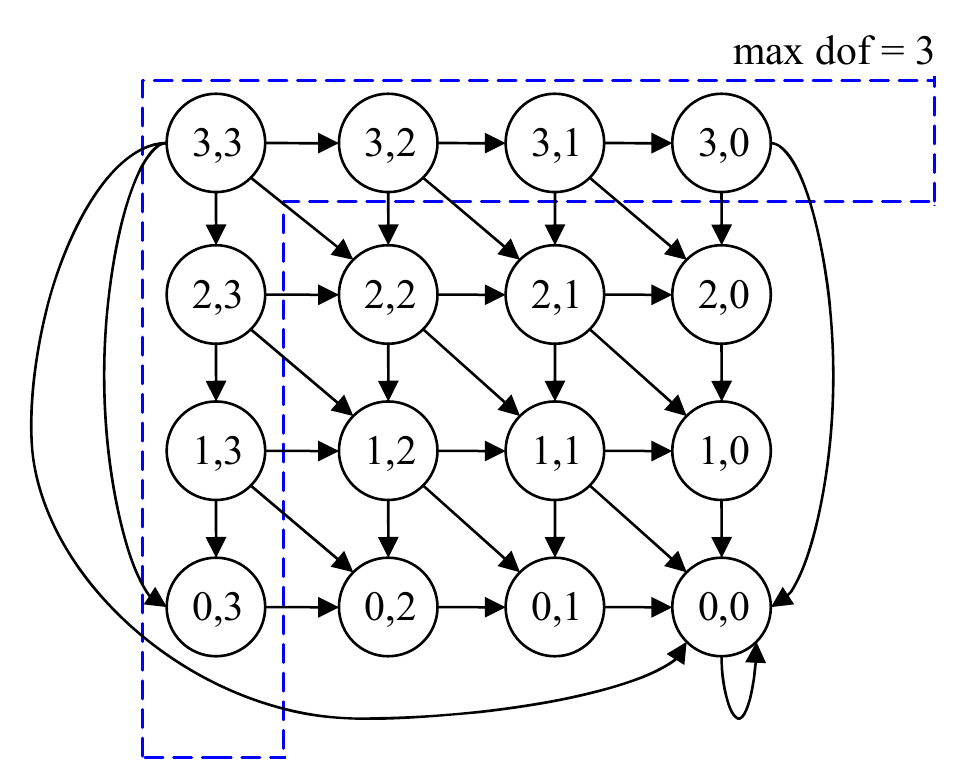}
\caption{Markov chain for the case of $N=2$ receivers and a block size of $M=3$.}
\label{MC2D.tag}
\end{figure}

	The process is modelled as a Markov chain. The states $(s_1,s_2,...,s_N)$ are defined by the number of dofs required $s_k$ at receiver $k$ to decode successfully the $M$ packets. Thus, the states range from $(M,M,...,M)$ to $(0,0,...,0)$. This is a Markov chain with ${(M+1)}^N - 1$ transient states and one recurrent state (state $(0,0,...,0)$). 
Figure~\ref{MC2D.tag} provides an example for 2 receivers and a block size of 3 packets. We have highlighted in this figure the states in which at least one receiver requires 3 coded packets in order to decode. Note that not all possible transitions from one state to the others have been included in this figure.  

The transition probabilities from state $(s_1,s_2,...,s_N)$ to state $(s_1',s_2',...,s_N')$ are
\begin{eqnarray} 
&P_{(s_1,s_2,...,s_N)\rightarrow (s_1',s_2',...,s_N')} = \notag\\&
P\left( \scriptstyle X_1(n) = s_1', ..., X_N(n) = s_N' | X_1(n-1) = s_1, ..., X_N(n-1) = s_N \displaystyle \right)
\end{eqnarray}
where $X_i(n)$ is the number of dof required at receiver $i$ at the end of transmission $n$. For simplicity of notation, let us say $P\left( \scriptstyle X_1(n) = s_1', ..., X_N(n) = s_N' | X_1(n-1) = s_1, ..., X_N(n-1) = s_N \displaystyle \right) = P\left( \scriptstyle s_1', ...,s_N' | s_1,... ,s_N \displaystyle \right)$. Similarly we consider that $P\left(  \scriptstyle X_i(n) = s_i'| X_1(n-1) = s_1, ..., X_N(n-1) = s_N \displaystyle \right) = P\left(  \scriptstyle s_i'| s_1, ..., s_N \displaystyle \right)$ and  $P\left(  \scriptstyle X_i(n) = s_i'| X_i(n-1) = s_i, \max _{j =1,2,...,N} s_j \displaystyle \right) = P\left(  \scriptstyle s_i'| s_i, \max _{j =1,2,...,N} s_j \displaystyle \right)$.

If we consider independent packet erasure channels for each of the receivers, 
\begin{eqnarray} 
P_{(s_1,...,s_N)\rightarrow (s_1',...,s_N')} = P\left(  \scriptstyle s_1'| s_1, ..., s_N \displaystyle \right)...P\left(  \scriptstyle s_N'| s_1, ..., s_N \displaystyle \right).
\end{eqnarray}

The dependence on the previous state $(s_1,s_2,...,s_N)$ can be translated into a dependence on the state with maximum dofs required to transmit, i.e. $i = \max _{j =1,2,...,N} s_j$, because $i$ determines $N_i$, the number of coded data packets sent by the transmitter. Thus,
\begin{eqnarray} 
&P_{(s_1,s_2,...,s_N)\rightarrow (s_1',s_2',...,s_N')} = \notag\\
&P\left(  \scriptstyle s_1'| s_1, \max _{j =1,2,...,N} s_j \displaystyle \right)...P\left(  \scriptstyle s_N'| s_N, \max _{j =1,2,...,N} s_j \displaystyle \right) =\notag\\
&P\left( \scriptstyle s_1'| s_1, N_i \displaystyle \right)...P\left(  \scriptstyle s_N'| s_N, N_i \displaystyle \right). \label{TransProb.tag}
\end{eqnarray}
where $P\left( \scriptstyle s_j'| s_j, N_i \displaystyle \right)$ has a similar structure to the transition probabilities studied in \cite{lucaniInfocom09}. The main difference is that the value of $N_i$ is no longer associated with the starting state of a particular receiver, but with a value determined from all starting states. For $0<s_j'< s_j$, this can be translated into
\begin{eqnarray}
&P\left( \scriptstyle s_j'| s_j, N_i \displaystyle \right) =\notag\\
&(1-Pe_{ack-j}) f(s_j,s_j') {(1-Pe_j)}^{s_j-s_j'} {Pe_j}^{N_{i} -s_j + s_j'}
\end{eqnarray}		
where 
\begin{align}
f(s_j,s_j') =
\begin{cases}
 \binom{N_{i}}{s_j - s_j'} & \text{if $N_{i} \geq s_j$,}
\\
0 & \text{otherwise}
\end{cases}
\end{align}
and $Pe_j$ and $Pe_{ack-j}$ represents the erasure probability of a coded packet and of an ACK packet for the erasure channel of receiver $j$, respectively.

For $s_j = s_j' > 0$ the expression for the transition probability reduces to:
\begin{eqnarray}
P\left( \scriptstyle s_j| s_j, N_i \displaystyle \right) = (1-Pe_{ack-j}){Pe_j}^{N_{i}} + Pe_{ack-j}.
\end{eqnarray}		 

Note that for $P\left( \scriptstyle 0| 0, N_i \displaystyle \right) = 1$. Finally, for $s_j' = 0$ $
P\left( \scriptstyle s_j' = 0| s_j, N_i \displaystyle \right) = 1 - \sum_{s_j' = 1} ^{s_j}  P\left( \scriptstyle s_j'| s_j, N_i \displaystyle \right)$.

We can define $P$ as the transition probability for our system. 
The speed of convergence from state $(M,...,M)$ to state $(0,...,0)$ can be summarized in the following lemma.

\textit{Lemma 1:} Let $\lambda_2$ be the second largest eigenvalue of $P$, and assume that there is only one eigenvalue with this magnitude. Then, the number of stops to wait for ACKs preceeded by transmissions of back-to-back coded packets $\aleph$ to transit from state $(M,...,M)$ to state $(0,...,0)$ with probability at least $1-\epsilon$ is
\begin{equation}
\aleph \geq \frac{\ln G - \ln \epsilon}{-\ln |\lambda_2|}
\end{equation}
where $G$ is a constant.

\begin{proof}
We use a similar method to the proof of lemma 2 in \cite{medard06}. 
Let us assume, without loss of generality, that our transition probability $P$ has the following structure
\begin{eqnarray}
P =
\left [ 
\begin{array}{cccc} 
P_{(M,...,M)\rightarrow (M,...,M)}& \cdot \cdot  && P_{(M,...,M)\rightarrow (0,...,0)} \\
:&:&:&:\\
0&\cdot \cdot & 0&  1\\
\end{array}  
\right] \notag
\end{eqnarray}
and 
\begin{eqnarray}
P^{N} =
\left [ 
\begin{array}{cccc} 
a_{11}(N)& a_{12}(N) &\cdot \cdot \cdot  & a_{1 \eta}(N) \\
:&:&:&:\\
0&\cdot \cdot & 0&  1\\
\end{array}  
\right] \notag
\end{eqnarray}
where $a_{ij}(N)$ is the probability of transitioning from the $i$-th state in the matrix to the $j$-th state in the matrix in $N$ iterations (transmissions followed by stop to receive ACKs), and $\eta$ are the number of columns in the matrix.

We are interested in determining the number of stops to wait for ACKs $\aleph$ so that $|a_{1 \eta}(N) - 1| \leq \epsilon$ with $\epsilon > 0$. In general, we could write this as $|q_0P^{\aleph} - \Pi| < T$, where $q_0$ is the starting state, $\Pi$ is the steady state probability, and $T$ is our performance target. In our case, $q_0 = [1, 0, ..., 0]$ since we are interested in studying convergence when we start in state $(M,...,M)$, $\Pi = [0, ..., 0, 1]$ because state $(0,...,0)$ is the only absorbing state, and $T = [T_1 ... T_{\eta - 1} \epsilon]$ where $\epsilon$ is our target performance, i.e. we have not imposed conditions for convergence from state $(M,...,M)$ to the other states ($T_1, ..., T_{\eta - 1}$). 

By the Cayley-Hamilton theorem, for $N \geq \eta$,
$P^N = \sum_{l = 0}^{\eta - 1} \phi_l(N) P^l $
for some constants $\phi_l(N)$. Denoting the eigenvalues by $1, \lambda_2,...,\lambda_{\eta} $, and using Lagrange's interpolation formula as in \cite{medard06}, we can write
\begin{equation}
P^N = \sum_{i=1}^{\eta} F_i^N (P) \lambda_i^N
\end{equation}
where
\begin{equation}
F_i^{N}(P) = \lambda_i^N \frac{\prod_{j = 1, j\neq i}^{\eta} (P - \lambda_jI)}{\prod_{j = 1, j\neq i}^{\eta} (\lambda_i - \lambda_j)}
\end{equation}
where $I$ is the identity matrix, and $\lambda_1 = 1$.

Since $P$ is a stochastic matrix, $|\lambda_i| < 1$ for $i>1$. Since $q_0 P^N \rightarrow \Pi$, as $N\rightarrow  \infty $, we have that $q_0 F_1^{\infty }(P) = q_0 F_1^{0 }(P) = q_0 F_1(P) = \Pi$. Thus, 
\begin{eqnarray}
|q_0P^N - \Pi | & \leq |\lambda_2|^N [1, 0, ..., 0] \sum_{i = 2}^{\eta} |F_i(P)| \\ 
&= |\lambda_2|^N [g_{11}, g_{12}, ...,  g_{1\eta}] 
\end{eqnarray}
where $F_i(P) = F_i^0(P)$, $|F_i(P)|$ denotes a matrix whose elements are the magnitudes of the elements of $F_i(P)$, and 
\begin{eqnarray}
\sum_{i = 2}^{\eta} |F_i(P)| =
\left [ 
\begin{array}{ccccc} 
g_{11}& \cdot \cdot \cdot  &  g_{1\eta} \\
:&:&:\\
g_{\eta 1}& \cdot \cdot \cdot & g_{\eta \eta} \\
\end{array}  
\right] \notag
\end{eqnarray}

Since we are interested in $|a_{1 \eta}(N) - 1| \leq \epsilon$, this translates to 
\begin{equation}
|\lambda_2|^{N} g_{1\eta} \leq \epsilon
\end{equation}
which concludes the proof.
\end{proof}

\itwsection{Mean Completion Time}

	The expected time for completing the transmission of the $M$ data packets constitutes the expected time of absorption, i.e. the time to reach state $(0,...,0)$ for the first time, given that the initial state is $(M,...,M)$. This can be expressed in terms of the expected time for completing the transmission given that the Markov chain is in state is $(s_1,...,s_N)$, $T_{(s_1,...,s_N)}$ , $\forall s_i = 0, 1, .. M - 1, \forall i = 1,...,N$. Let us denote the transmission time of a coded packet as $T_{p}$, and the waiting time to receive an ACK packet as $T_{w}$. For our scheme, $T_{p} = \frac{h +n+gM}{R}$, as in \cite{lucaniInfocom09}, but the expression of $T_w$ changes slightly to consider the transmission of multiple ACK packets from the receivers to the transmitter. 

Let us define $d_i$ as the distance between the transmitter and node $i$, as in Figure~\ref{BroadcastNetwork.tag}. We assume that the nodes have been numbered so that $d_1 \leq d_2 \leq ... \leq d_N$. Let us define $t_{btA}^i$ as the time node $i$ has to wait before starting to transmit after he has received the last coded packet from the transmitter. The choice of $t_{btA}^i$ depends on characteristics of the link between the receivers and the transmitter and interference that a receiver could generate in other receivers at the time of transmitting its ACK. If the nodes do not generate interference over other nodes, e.g. a satellite link which typically has a very directional antennas, then we could use $t_{btA}^i = \max \left( t_{btA}^{i-1} + T_{ack} - T_{rt-i} + T_{rt-(i-1)}, 0 \right) $, where $T_{ack} = n_{ack}/R$, $n_{ack}$ is the number of bits in the ACK packet, $R$ is the link data rate, and $T_{rt-i}$ is the round trip time for node $i$. Thus, $T_{w} = T_{rt-N} + t_{btA}^N + T_{ack}$ and $t_{btA}^1 = 0$. 
If the transmission of the ACK packets can create interference in transmissions to other receivers, the first ACK could be sent after all data packets have been correctly received. In this case, $t_{btA}^1 = \left(T_{rt-N} - T_{rt-1}\right)/2$ and we can use the previous recursive formula for $t_{btA}^i$ and the expression for $T_{w}$. 

We can define $T^i$ as the time it takes to transmit $N_i$ coded data packets and receive the ACK packets from the different receivers. It is easy to show that $T^i = N_i T_{p} + T_{rt-N} + t_{btA}^N + T_{ack}$.

The mean completion time when the system is in state $(s_1,...,s_N)$ is given by
\begin{eqnarray}
&T_{(s_1,...,s_N)} = T^i +\\& \sum_{(s_1,...,s_N),(s_1',...,s_N')} P_{(s_1,...,s_N) \rightarrow (s_1',...,s_N')} T_{(s_1',...,s_N')} 
\end{eqnarray}
where $i = \max_{j = 1,...,N} s_j$.
We can express this in vector form as
\begin{equation}
\bar T = {\left[ I - P \right]}^{-1} \bar \mu. 
\end{equation} 
where $\bar T = [ T_{(s_1,...,s_N)} ]$, $\bar \mu = [ T^i]$ and $P$ is the corresponding transition probability.

Since we are interested in the mean completion time when we start at state $(M,...,M)$, we can use Cramer's rule to determine
\begin{equation}
T_{(M,...,M)} = \frac{\det\left( \Gamma \leftarrow_{(M,...,M)} \bar \mu \right) }{\det \left( \Gamma \right)}
\end{equation}
where $\Gamma = I - P$, and the notation $\Gamma \leftarrow_{(M,...,M)} \bar \mu$ represents a matrix that has all columns as the $\Gamma$ matrix except the column corresponding to state $(M,...,M)$ which is substituted by the vector $\bar \mu$. 
Due to characteristics of the Markov chain, $\Gamma$ is a triangular matrix. Thus, computing $\det \left( \Gamma \right)$ reduces to multiplying the elements in the main diagonal of the $\Gamma$ matrix.

The expected time for each state depends on all the expected times for the previous states. However, optimizing the values of all $N_{i}$ is not as straight forward as the recursive method used in \cite{lucaniInfocom09}. 

Also, note that there are ${(M+1)}^N$ states in our Markov chain. This implies that we have to compute the transition probabilities to fill a $\left({(M+1)}^N -1 \right)$~x~$\left({(M+1)}^N -1 \right)$ matrix and then solve the determinants of matrices of the same dimensions for each iteration of a search algorithm. Thus, the computational demands increases significantly, specially when we increase the number of receivers. 

Then, let us consider some heuristics to estimate the values of $N_{i}, \forall i = 1,...,M$, either to use them directly as an approximate solution or as an initial point of a search algorithm. These heuristics rely on solving the link case \cite{lucaniInfocom09} considering as packet erasure probability of the link a function of the packet erasure probabilities of the different channels in broadcast.
 
\textit{1) Worst Link Channel:} In this heuristic we approximate the system as a link to the receiver with the worst channel, i.e. $Pe = \max_j Pe_j$. Then, we compute $N_{i}, \forall i = 1,...,M$ to minimize the mean completion time as in \cite{lucaniInfocom09} using the current values of $T_p$, $T_w$, and $Pe_{ack} = \max_j Pe_{ack-j}$.

\textit{2) Combined Erasure Effect:} In this heuristic we approximate the system as a link to a receiver with $Pe = 1 - \prod_j(1 - Pe_j)$, i.e. assuming that a coded packet suffers an erasure in the link when it is seen as an erasure by at least one receiver. Then, we compute $N_{i}, \forall i = 1,...,M$ to minimize the mean completion time as in \cite{lucaniInfocom09} using the current values of $T_p$, $T_w$, and $Pe_{ack} = 1 - \prod_j(1 - Pe_{ack-j}) $.  


Determining $P$ for a link requires computing $O\left( M^2 \right)$ transition probabilities, while solving the same problem for broadcast requires a computation of $O\left( {\left( N(M+1)\right)}^{2N}  \right)$ equivalent transition probabilities. 
This does not include the savings provided by inverting considerably smaller matrices. 

Also, note the first heuristic is optimistic, disregarding the effect of nodes with better channels, while the second heuristic is pessimistic, concentrating the effect of all losses in one link. Since the $N_i$'s increase as the probability of erasure increases, the 'Worst Link Channel' and 'Combined Erasure Effect' heuristics provide a lower and upper bound on the values of $N_i, \forall i$, respectively.  

Finally, it is important to emphasize that the $N_i$'s do not need to be computed in real time. As explained in \cite{lucaniInfocom09}, they can be pre-computed and stored in the receiver as look-up tables to reduce the computational load on the nodes. The nodes only have to choose the appropriate $N_i$'s from the tables, considering channel conditions at the time of transmission. 

\itwsection{Comparison Schemes}
In this section, we extend the work in \cite{Atilla07} to determine the mean completion time for optimal scheduling policies for broadcast. These policies consider no coding of the data packets, no channel state information, and nodes that only ACK when they have received all $M$ data packets. As in \cite{Atilla07}, we restrict the analysis to independent symmetric channels, i.e. $Pe_1 =...=Pe_N$, and no erasures in the ACKs for tractability. Note that we had no such restrictions in our network coding scheme. Our contribution includes 1) considering the effect of $T_{rt}$, $T_p$, and $T_{ack}$, and 2) the characterization for full duplex and TDD channels.

\textit{1) Broadcast with Round Robin in Full Duplex Channel (RR Full Duplex):} The objective is to transmit $M$ data packets to all users. Since the channels are independent and identically distributed over time and users, one of the optimal policies is Round Robin (RR). Thus, packet $k$ in the block is transmitted every $(mM + k)T_p$ time units for $m=0,1,2,...$ until all the receivers get all $M$ packets \cite{Atilla07}. Using a similar analysis as in \cite{Atilla07},
\begin{equation}
E[T] = T_{w} + T_p M \left( \gamma + E[\max_{i,k}X^i_k] \right)
\end{equation}
where $1 + X^i_k$ is the number of transmissions of packet $k$ needed to reach node $i$, $\gamma \in (1/2, 1)$, and
\begin{equation}
E[\max_{i,k}X^i_k] = \sum_{t=1}^{\infty } \left[ 1 - {(1-Pe^t)}^{MN}\right]
\end{equation}
where $Pe = Pe_1 =...=Pe_N$. Note that $\gamma = 1$ and $\gamma = 1/2$ give us an upper and lower bound on the mean completion time, respectively.

\textit{2) Broadcast with Round Robin in TDD (RR TDD):} This scheme assumes limited feedback due to the TDD constraint. We assume that the transmitter broadcasts all $M$ packets back-to-back, then stops to receive ACK packets that indicate completion of the entire file. If there are nodes that have not acknowledged the block of packets, the transmitter repeats the process, i.e. sends all $M$ packets and stops to listen for ACKs. We can express the mean completion time of this scheme as
\begin{equation}
E[T] =  \left( T_{w} + T_p M \right) E[\max_{i,k}X^i_k].
\end{equation}



\itwsection{Numerical Results} 

\begin{figure}[t]
\centering	
\includegraphics[height=3.5in,width=3.5in]{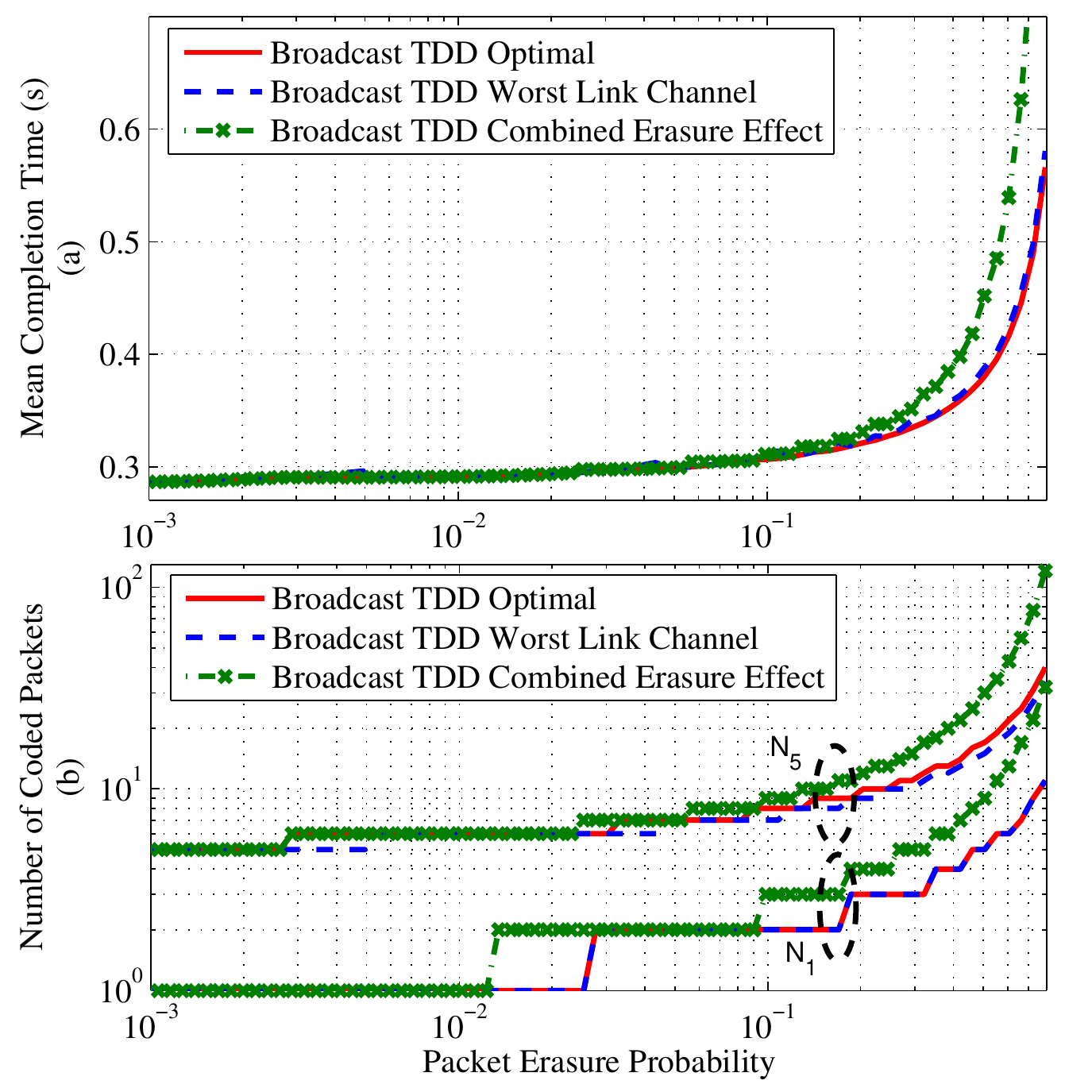}
\caption{(a)Mean completion time and (b) number of coded packets $N_5$ and $N_1$, for the optimal choice of $N_i$'s and two heuristics, for $N=2$ receivers at the same distance from the transmitter, $M=5$,  packet erasure probability is the value for the two independent channels, $R=1.5$~Mbps, $h = 80$~bits, $g = 20$~bits, $n_{ack} = 50$~bits}
\label{Broadcast2_M5_ChangetogetherPe.tag}
\end{figure}    

This section provides numerical examples that compare the performance of our network coding scheme for broadcast in TDD channels, considering a satellite example. In particular, we compare the performance of the scheme when the $N_i$'s are 1)chosen to minimize the mean completion time, 2)chosen using the 'Worst Link Channel' heuristic, and 3) chosen using the 'Combined Erasure Effect' heuristic. The comparison is carried out in terms of the mean completion time of $M$ data packets under different packet erasure probabilities. We show that the 'Worst Link Channel' provides close-to-optimal performance with the advantage of reducing the computational load on the search algorithm. For simplicity, we consider that there are no erasures of ACK packets and that the distance between the transmitter and each receiver is the same. The latter is a good approximation in many satellite scenarios. Finally, we compare our broadcast scheme with RR TDD and RR Full Duplex.

\begin{figure}[t]
\centering	
\includegraphics[height=3.3in,width=3.3in,keepaspectratio]{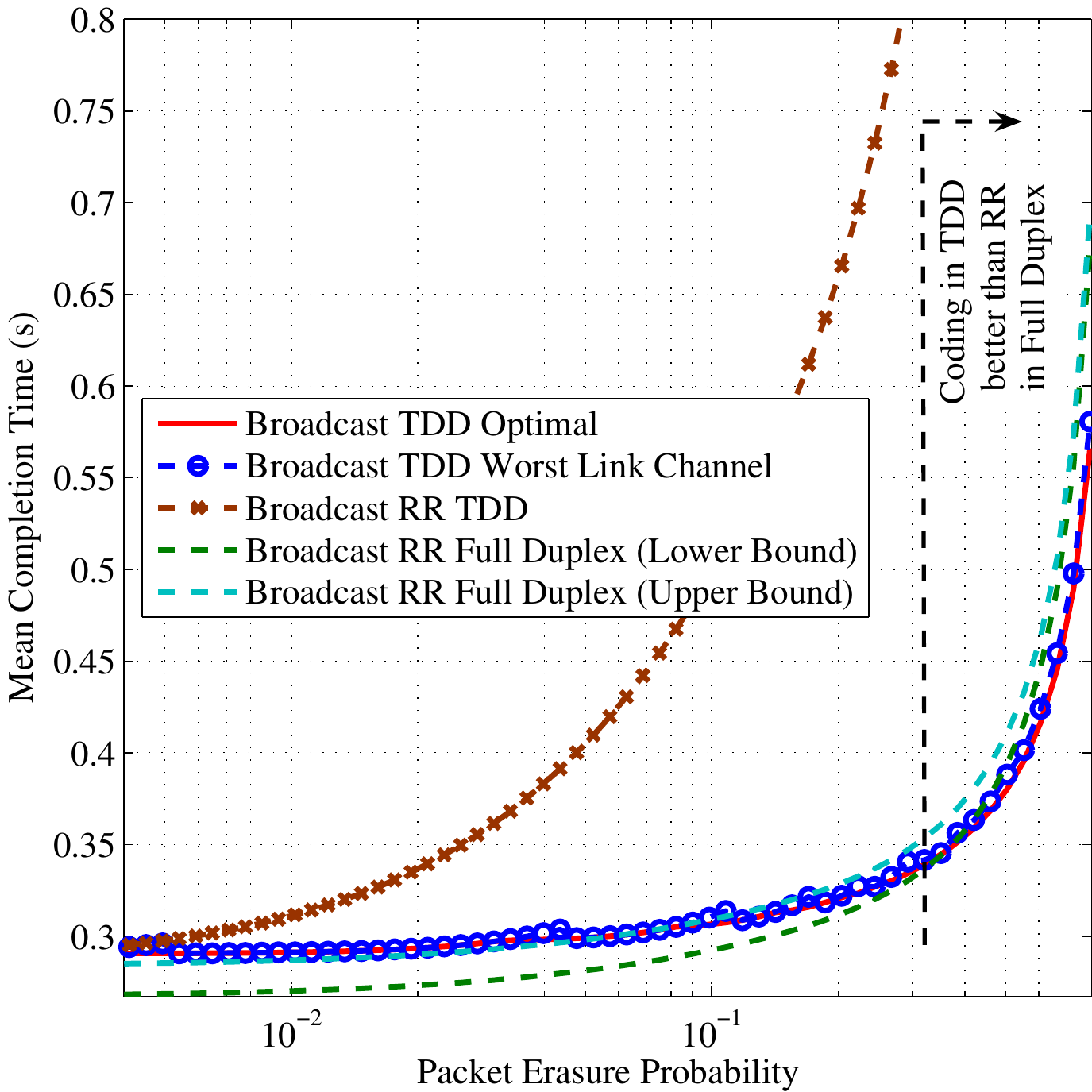}
\caption{Mean completion time for the optimal choice of $N_i$'s, 'Worst Link Channel' heuristic, and Round Robin Broadcast schemes with Full duplex and TDD channels. We use as parameters $N=2$ receivers at the same distance from the transmitter, $M=5$, packet erasure probability is the value for the two independent channels, $R=1.5$~Mbps, $h = 80$~bits, $g = 20$~bits, $n_{ack} = 50$~bits}
\label{Broadcast2_M2_BetterthanScheduling.tag}
\end{figure}


Figure~\ref{Broadcast2_M5_ChangetogetherPe.tag} shows (a) the mean completion time and (b)number of coded packets $N_5$ and $N_1$, for the optimal choice of $N_i$'s and our two heuristics when we have independent channels with a common packet erasure probability, i.e. $Pe_1 = Pe_2$. We consider data packets of size $n = 10,000$ bits in a GEO satellite link with a propagation delay of 125~ms, and the parameters specified in the figure. 

Figure~\ref{Broadcast2_M5_ChangetogetherPe.tag}(a) illustrates that choosing $N_i$'s using the 'Worst Link Channel' heuristic provides close-to-optimal performance in terms of mean completion time for a wide range of packet erasure probabilities. Although, the 'Combined Erasure Effect' heuristic provides a better estimate for low packet erasure probabilities in this case, its choice of $N_i$'s for high packet erasures produces considerably higher completion times. This fact is explained because the 'Combined Erasure Effect' heuristic is pessimistic in terms of the amount of coded packets that are successfully received.       

Figure~\ref{Broadcast2_M5_ChangetogetherPe.tag}(b) shows that the optimal choice of $N_i$'s is bounded by the choices of $N_i$'s using our two heuristics. As explained in Section III, this is not surprising because one of the heuristics is optimistic in its approximation ('Worst Link Channel') and the other is pessimistic ('Combined Erasure Effect'). This result is interesting at the time of developing an algorithm to search for the optimal value, because we could limit the search to the values given by the heuristics.

Since the choice of $N_i$'s using the 'Worst Link Channel' heuristic provides a performance that is close to the optimal, we could use it as an initial choice of the $N_i$'s so that a search algorithm finds the optimal $N_i$'s, or we could use them directly.  
However, for large values of $N$ and $M$ a full search procedure might become exceedingly expensive in terms of computation time. Computing the $N_i$'s using the 'Worst Link Channel' heuristic is easily performed even for large values of $M$, because it approximates the system as a link. Reference \cite{lucaniInfocom09} shows some examples for cases of $M = 90$ and $M=130$. In practice, using the heuristic provides a good trade-off between complexity and accuracy.
     
Figure~\ref{Broadcast2_M2_BetterthanScheduling.tag} compares the performance of our Broadcast TDD scheme with $N_i$'s computed optimally and with the 'Worst Link Channel' heuristic, and compares it to the performance of RR TDD and RR Full Duplex. First, note that for the range of packet erasures considered, our coding scheme performs at least as good as the RR TDD, and considerably better at high erasures. Second, the performance of our scheme is very close to that of the RR Full Duplex for low erasures. However, our coding scheme performs better at high packet erasures ($Pe_1 = Pe_2 > 0.3$), e.g. at $Pe_1 = Pe_2  =0.8$ the RR Full Duplex scheme takes 20\% more time to complete transmissions. Thus, even with a single channel for data and feedback, i.e. half of the resources, we can perform better by tailoring coding and feedback appropriately.     
     
           

\itwsection{Conclusion}

This paper provides an extension to the use of random linear network coding over channels where time division duplexing is necessary. In particular, we study the case of broadcasting a block of $M$ data packets to $N$ receivers. Similar to our work in \cite{lucaniInfocom09} and \cite{lucaniICC09}, the scheme considers that a number of coded data packets are transmitted back-to-back before stopping to wait for the receivers to acknowledge how many degrees of freedom, if any, are required to decode the information correctly. 

We prove that the number of stops to listen for the ACK packets $\aleph$ in order to complete transmission with probability at least $1-\epsilon$, for any $\epsilon >0$, is $\aleph \geq \frac{\ln G - \ln \epsilon}{-\ln |\lambda_2|}$. Here we considered $\lambda_2$ be the second largest eigenvalue of the transition probability matrix $P$, assumed that there is only one eigenvalue with this magnitude, and $G$ is a constant.
	

We also provide a simple heuristic to compute the number of coded packets to be sent before stopping that achieves close to optimal performance with the advantage of a considerable reduction in the search time. This heuristic approximates the system as a link to the receiver with the worst channel, and computes the number of coded packets to minimize completion time as in \cite{lucaniInfocom09}. This heuristic provides a good trade-off between computational complexity and performance.

Numerical results show that our coding scheme outperforms a Round Robin broadcast scheme in an TDD channel. More importantly, for high packet erasures, our coding scheme for TDD outperforms a RR scheme operating in a full duplex channel.

	Future research will extend the problem of broadcast to cases in which nodes forming a cluster are allowed to cooperate and share dofs in order to decode the information. Also, we will consider extensions to the general problem of wireless networks.


\itwacknowledgments

This work was supported in part by the National Science Foundation under grants No. 0520075, 0831728 and CNS-0627021, by ONR MURI Grant No. N00014-07-1-0738, subcontract \# 060786 issued by BAE Systems National Security
Solutions, Inc. and supported by the Defense Advanced Research Projects
Agency (DARPA) and the Space and Naval Warfare System Center (SPAWARSYSCEN),
San Diego under Contract No. N66001-06-C-2020 (CBMANET), subcontract \# 18870740-37362-C
issued by Stanford University and supported by the DARPA.

\end{itwpaper}

\begin{itwreferences}


\bibitem{lucaniInfocom09}Lucani, D. E., Stojanovic, M., M\'{e}dard, M., ``Random Linear Network Coding For Time Division Duplexing: When To Stop Talking And Start Listening", to appear in INFOCOM'09, available at arXiv:0809.2350v1[cs.IT]

\bibitem{lucaniICC09}Lucani, D. E., Stojanovic, M., M\'{e}dard, M., ``Random Linear Network Coding For Time Division Duplexing: Energy Analysis", to appear in ICC'09, available at arXiv:0901.0269v1[cs.IT]

\bibitem{ho06}
Ho, T., Medard, M., Koetter, R., Karger, D.R., Effros, M., Shi, J., Leong, B.,``A Random Linear Network Coding Approach to Multicast", Trans. Info. Theory, vol. 52, no. 10, pp.4413-4430, Oct. 2006

\bibitem{medard06} M\'{e}dard, M., Srikant, R., ``Capacity of Nearly Decomposable Markovian Fading Channels Under Asymmetric Receiver-Sender Side Information", IEEE Trans. on Info. Theory, vol. 52, no. 7, pp. 3052-3062, Jul. 2006

\bibitem{Atilla07}Eryilmaz, A., Ozdaglar, A., M\'{e}dard, M., ``On Delay Performance Gains from Network Coding", In Proc. CISS'06, pp. 864-870, Princeton, NJ, USA, Mar. 2006

\end{itwreferences}

\end{document}